\documentclass[conference]{IEEEtran}
\usepackage{amsthm,amssymb,graphicx,multirow,amsmath,color,amsfonts}
\usepackage[update,prepend]{epstopdf}
\usepackage[noadjust]{cite}
\usepackage[latin1]{inputenc}
\usepackage{tikz} 
\usepackage{bbm} 
\usepackage{pdfpages}
\usepackage{multirow}
\usepackage{comment}

\usepackage{algorithm}
\usepackage{algpseudocode}

\usepackage{balance}

\usepackage{subfig}
\usepackage{mathrsfs}
\usepackage[normalem]{ulem}
\usepackage{optidef}
\allowdisplaybreaks 
\usepackage[colorinlistoftodos,bordercolor=orange,backgroundcolor=orange!20,linecolor=orange,textsize=scriptsize]{todonotes}

\usepackage{setspace}	

\graphicspath{{figures/}}

\newcommand{\algorihmicserver}{\textbf{Global Server}}
\algdef{SE}[SERVER]{Server}{EndServer}[1]{\algorihmicserver\ #1\ \algorihmicdo}{\algorihmicend\ \algorihmicserver}%
\algtext*{EndServer}
\newcommand{\algorihmicclient}{\textbf{Client k}}
\algdef{SE}[CLIENT]{Client}{EndClient}[1]{\algorihmicclient\ #1\ \algorihmicdo}{\algorihmicend\ \algorihmicclient}%
\algtext*{EndClient}
\makeatletter
\newcommand{\brokenline}[2][t]{\parbox[#1]{\dimexpr\linewidth-\ALG@thistlm}{\strut\raggedright #2\strut}}
\makeatother

\usepackage{romanbar}

\makeatletter
\renewcommand{\fnum@figure}{Figure \thefigure}
\makeatother
\usepackage{tabularx}
\makeatletter
\renewcommand{\fnum@table}{Table \thetable}
\makeatother

\begin{document}
\include{notation}
\title{
Semantic-Aware Resource Allocation in Constrained Networks with Limited User Participation 
}

\author{
    \IEEEauthorblockN{Ouiame Marnissi, Hajar EL Hammouti, El Houcine Bergou
        }
    \IEEEauthorblockA{ College of Computing, Mohammed VI Polytechnic University, Ben Guerir, Morocco.\\
\{ouiame.marnissi, hajar.elhammouti, elhoucine.bergou\}@um6p.ma
}
}
\maketitle

\begin{abstract}


Semantic communication has gained attention as a key enabler for intelligent and context-aware communication. However, one of the key challenges of semantic communications
is the need to tailor the resource allocation to meet the
specific requirements of semantic transmission. In this paper, we focus on networks with limited resources where devices are
constrained to transmit with limited bandwidth and power over
large distance. Specifically, we devise an efficient strategy to select the most pertinent semantic features and participating users, taking into account the channel quality, the transmission time, and the recovery accuracy. To this end, we formulate an optimization problem with the goal of selecting the most relevant and accurate semantic features
over devices while satisfying constraints on transmission
time and quality of the channel. This involves optimizing communication resources, identifying participating users, and choosing specific semantic information for transmission. The underlying problem is inherently complex due to its non-convex nature and combinatorial constraints. To overcome this challenge, we efficiently approximate the optimal solution by solving a series of integer linear programming problems.
Our numerical findings illustrate the effectiveness and efficiency of our approach in managing semantic communications in networks with limited resources.


\end{abstract}
\section{Introduction} \label{sec:intro}

In the last few years, semantic communication has gained attention as a key paradigm that enables intelligent and context-aware communication. Unlike conventional communication technologies \cite{marnissi}, where bit sequences are transmitted regardless the context of the communication, semantic communication conveys the meaning of transmitted data in accordance with the communication's context, and transmits a concise representation of the inferred message. As a consequence, it introduces more intelligence into communication systems, allowing them to make more informed decisions. Additionally, it enhances the communication reliability by enabling the receiver to reconstruct missing segments of a corrupted message based on contextual semantics. 
 Most importantly, it enhances the communication efficiency by reducing the amount of transmitted data.

One of the key challenges of semantic communications is the need to tailor the resource allocation to meet the specific requirements of a semantic transmission. In contrast to traditional wireless communications, where performance metrics are well-defined and the resource allocation model is well-established, the research community is still working on identifying the most appropriate metrics and resource allocation models to improve the performance of a semantic system. In this context, our paper formulates the problem of resource allocation for semantic communication, taking into account factors like transmission time, relevance of transmitted semantic features, and their accuracy at the recovery. It addresses the question: How can we efficiently select the most relevant semantic features in a resource-constrained network with limited user participation?


Semantic communication has been considered for various applications including text 
\cite{lite2021deepl,ResAlloc_2022,attention2022saad}, speech \cite{speech2021}, and image transmissions \cite{Img2022zhang}. To assess the performance of semantic systems for such applications, new metrics have been introduced. In \cite{KG2022entrop}, the authors proposed a novel metric known as the semantic similarity score. This score is determined by converting text into vectors and evaluating the degree of similarity between these vectors. In \cite{speech2021}, another approach is proposed to measure the similarity of the original and recovered speech transcription by defining character-error-rate and word-error-rate in terms of character/word substitutions, character/word deletions and character/word insertions. To compare the performance of semantic and conventional communication, authors in \cite{Theory2023semco} provided a definition of a universal semantic entropy which describes the uncertainty in the semantic interpretation of received symbols. Based on this definition, they show that the transmission rate of semantic communication can exceed the Shannon channel capacity.

These performance metrics have been introduced to enhance resource management in the context of semantic communications. In this regard, the authors in \cite{imran2022} have devised a strategy that combines user selection and bandwidth allocation, aiming to maximize the semantic throughput. A related metric, as well as semantic efficiency, are explored in \cite{ResAlloc_2022}, where an efficient resource allocation approach is proposed. This method leverages the DeepSC tool for text transmission. Text-based semantic communication is also investigated in \cite{attention2022saad}, with a joint resource block allocation and semantic information selection. The optimization problem in this work focused on maximizing the semantic similarity between the original text and the received semantic information while adhering to latency constraints.

In this paper, we focus on networks with limited resources, such as Internet of Things (IoT) networks, where
devices are constrained to transmit with limited bandwidth and power over large distances. In this context,
a semantic filtering module is incorporated to select a subset of semantic features along with the users participating in the communication. Our focus is to devise a strategy to choose the
most relevant semantic features considering criteria such as the
quality of the channel, the transmission time, and the accuracy
of recovery. Our contributions can be summarized as follows
\begin{itemize}
    \item We formulate an optimization problem that aims to maximize the most relevant and accurate semantic features over devices while satisfying constraints on transmission time and quality of the channel. 
    \item The underlying optimization problem falls
into the category of mixed integer non-linear programming (MINLP). To solve this problem, we first address the optimization of transmit power and bandwidth. The problem then reduces into an optimization of semantic features and device selection which we address using an alternating approach that solves at each iteration a linear programming problem with the guarantee to converge  to an optimum. We provide also the worst case complexity (WCC) bounds of our algorithm.
\item Simulation results show the performance of our approach
compared to other semantic-aware resource allocations.
\end{itemize}

The remainder of this paper is organized as follows. The system model is described in section II. In section III, we formulate the studied problem as a joint optimization of power, bandwidth, semantic extraction and user selection. Our proposed approach is described in section IV. Simulation results are provided and analyzed in section V. Finally, section VI draws the conclusions of our paper.
\section{System Model}
\begin{figure}[htbp]
\centerline{\includegraphics[scale = 0.45]{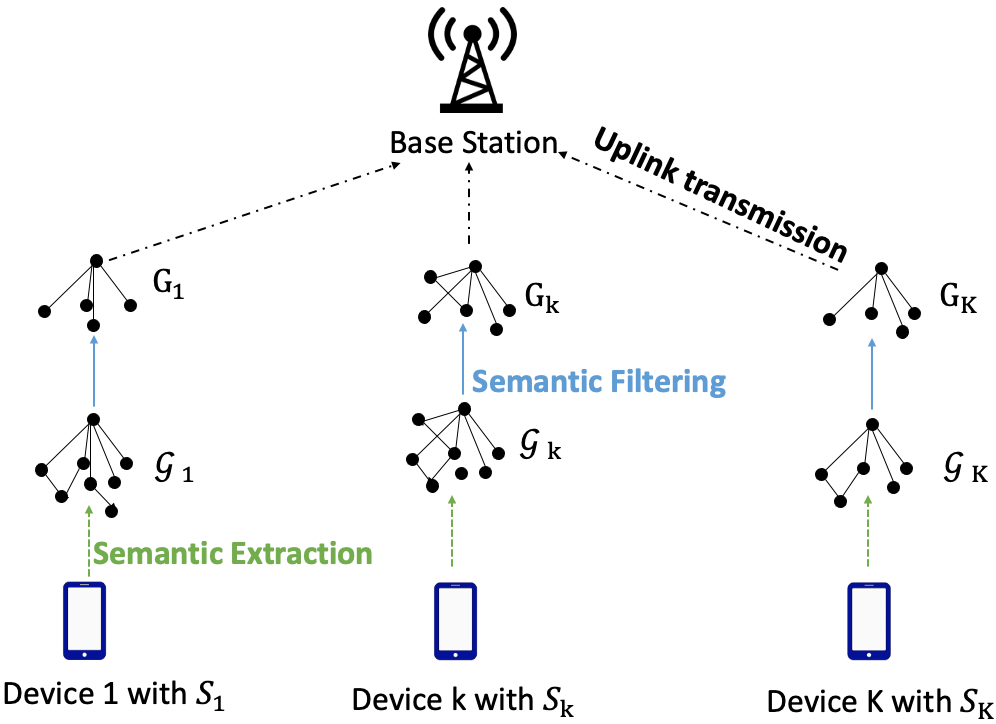}}
\caption{Semantic Communication Model}
\label{SemCom_model}
\end{figure}

Consider a wireless communication system where a set of devices communicate with a server located at the cloud and accessed through a base station (BS) as shown in Figure \ref{SemCom_model}. We suppose that devices collect data locally and transmit relevant information to the server which performs some inference tasks. 
We denote by $\mathcal{K}=\{1,\dots,K\}$ the set of devices. Let $\boldsymbol{s}_k$ be the source data of device $k$. $\boldsymbol{s}_k$ is passed through a semantic extraction module, which outputs a compact representation of the data. In this paper, we adopt a graph representation of the semantic information. An effective way to represent the extracted information is through an entity-relation graph. In this graph representation, the entities extracted from the data are depicted as nodes, while the relationships among them are represented as edges connecting the corresponding nodes\footnote{Other graph representations can be used such as \textit{concept graph} where concepts extracted from the data are represented as nodes, and their relationships, such as synonymy and antonym are represented as edges. }. To enable such an extraction and tailor it to the desired task, we suppose that the server shares with each device $k$ a \textit{knowledge base}, represented by a vector $\boldsymbol{\mathcal{C}_k}$. The knowledge base provides a contextual understanding to the semantic extraction and helps identify which information is relevant to the task. Let $\boldsymbol{\mathcal{G}}_k$ be the graph representation of the semantic information of device $k$ obtained via the semantic extraction module and the knowledge base $\boldsymbol{\mathcal{C}_k}$. We suppose that the graph $\boldsymbol{\mathcal{G}}_k$  is composed of $N_k$ triplets in the form of entity-relation-entity. We denote the $n-th$ triplet of device $k$ by $\boldsymbol{\mathcal{G}}_k^n$.

In general, the extracted semantic information, represented by a graph, is encoded into bits and transmitted to the server via the wireless channel. 
However, in networks with limited resources, such as Internet of Things (IoT) networks, where devices are  constrained to transmit with limited bandwidth and power over large distances, the size of the transmitted data needs to be further reduced. Hence, to achieve this purpose, a semantic filtering module is incorporated to select a subset $\boldsymbol{G}_k$ of the semantic graph $\boldsymbol{\mathcal{G}}_k$.  This subset is then mapped into bits and sent to the server through the wireless channel. At the reception, the server proceeds with channel and semantic decoding to recover the transmitted semantic features. We denote by $\hat{\boldsymbol{G}}_k$ the recovered semantic features. We suppose that $G_k = \boldsymbol{\eta_k} {\mathcal{G}}_k$ where $\boldsymbol{\eta_k}= [\eta_{k}^1,...,\eta_{k}^n,...,\eta_{k}^{N_k}]$ is the triplets assignment vector of the $k$-th user with $\eta_{k}^n \in \{0,1\}$. Here $\eta_{k}^n = 1$ when the $n$-th triplet of user $k$ is chosen for transmission, and $\eta_{k}^n = 0$, otherwise. 

\subsection{Importance Score}

To quantify the level of relevance between a specific sub-graph and the target task, we define the \textit{importance score} as a similarity metric that measures the proximity between the selected subset and the knowledge base. The importance score, similar to the similarity measure used to evaluate the resemblance between two sentences, quantifies the level of importance or relevance between each element of the graph and the knowledge base. Accordingly, the importance score $I_k^n$ between the triplet $\boldsymbol{\mathcal{G}}_k^n$ and the knowledge base $\mathcal{C}_k$ is given by
\begin{equation}  
I_k^n=\frac{\boldsymbol{V}(\boldsymbol{\mathcal{G}}_k^n)\boldsymbol{V}(\boldsymbol{\mathcal{C}_k})^T}{||\boldsymbol{V}(\boldsymbol{\mathcal{G}}_k^n)||\times||\boldsymbol{V}(\boldsymbol{\mathcal{C}_k})||},
\end{equation}
where $V(.)$ is the Bidirectional Encoder Representations from Transformers (BERT) model, which is a pre-trained language model introduced by Google in 2018 \cite{BERT2019}. The importance score ranges between $0$ and $1$, where a higher value indicates a greater degree of relevance to the desired task.  

While the relevance of transmitted data is crucial for a successful semantic communication in resource-limited networks,  other factors need to be taken into account during the transmission including the channel impact. In the following, we describe how we model the wireless communication between devices and the BS and account for it for a reliable semantic communication. 



\subsection{Channel Impact}


We suppose that devices use orthogonal frequency domain multiple
access (OFDMA) for their exchange with the BS. Let $P_k$ be the transmit power of device $k$, $d_k$ the distance between the device and the BS, and $h_k$ the fading channel response assumed to be Rayleigh. The signal-to-noise-ratio (SNR) related to device $k$, $\gamma_k$ is given by 

\begin{equation}
\begin{split}
        \gamma_k(P_k)&=\frac{P_kd_k^{-2}|h_k|^2}{\sigma^2}\\
        &=P_kD_K,
\end{split}
\end{equation}
where $\sigma^2$ is the variance of an additive white Gaussian noise and $D_k = \frac{d_k^{-2}|h_k|^2}{\sigma^2}$ for simplification.

Let $B$ be the total available network bandwidth that is shared between multiple devices. We denote by $0 \leq b_k \leq 1$ the proportion  of the total bandwidth allocated to device $k$. Accordingly, the data rate of device $k$ can be written 

\begin{equation}
r_k(b_k,P_k)=b_k \times B \log_2(1+\gamma_k(P_k)).
\end{equation}
 Let $S_k^n$ be the corresponding size in bits of the triplet $\boldsymbol{\mathcal{G}}_k^n$. The required time to transmit $\boldsymbol{G}_k$ is given by 

 \begin{equation}     T_k(\eta_k,b_k,P_k)=\frac{\sum_{n=1}^{N}\eta_k^nS_k^n}{r_k(b_k,P_k)}.
 \end{equation}

While a higher data rate allows for the transmission of more bits in a short time frame, it can also lead to a large number of errors, particularly in scenarios with limited unfavorable channel quality. Consequently, simply increasing the data rate without addressing the bit-error-rate (BER) may result in many errors, potentially leading to wrong semantic decoding. Hence, it is essential to ensure that the BER is below a specific threshold. To this end, we introduce a general function that characterizes the relationship between the BER and the SNR. Let us denote this function as $f$, which is assumed to be a decreasing function of the SNR. The specific form of the function $f$ varies depending on the modulation scheme. The BER related to device $k$ should be below a threshold $\beta^{\rm th}$, i.e.,

\begin{equation}
f(\gamma_k(P_k))\leq \beta^{\rm th}.
\end{equation}

To achieve a successful semantic communication, it is important to consider two key factors: minimizing the transmission time and ensuring an appropriate BER. In fact, in semantic communication systems, which often involve interactive exchanges between devices and servers, prompt responses are essential for timely task execution. Additionally, a high BER can introduce substantial errors during the recovery of semantic features, potentially compromising the accuracy of the semantic communication.

In the following subsection, we outline the final step of a semantic transmission, which involves the recovery of semantic features.

\subsection{Semantic Recovery}
In addition to errors introduced by the communication channel, the semantic decoding may be subject to other errors due to two main sources. First, ambiguity may arise when a semantic feature have multiple meanings, leading to a misinterpretation by the receiver. Second, semantic distortion can occur due to a misalignment of the knowledge base between the transmitter and the receiver. To evaluate the semantic mismatch, we use the cosine similarity metric, $R_k^n$, to measure the similarity between the original element $\boldsymbol{\mathcal{G}}_k^n$ and the recovered one $\hat{\boldsymbol{\mathcal{G}}_k^n}$, given by

\begin{equation}  R_k^n=\frac{\boldsymbol{V}(\boldsymbol{\mathcal{G}}_k^n)V(\hat{\boldsymbol{\mathcal{G}}_k^n})^T}{||\boldsymbol{V}(\boldsymbol{\mathcal{G}}_k^n)||\times||\boldsymbol{V}(\hat{\boldsymbol{\mathcal{G}}_k^n})||}.
\end{equation}

To ensure a successful semantic communication in resource-constrained networks, we aim to select the most relevant information with the highest semantic recovery. To achieve this goal, we introduce a semantic efficiency metric which takes into account both the importance score of a selected triplet and its semantic recovery. Accordingly, the semantic efficiency (SE) metric of device $k$ with a triplet selection vector $\boldsymbol{\eta_{k}}$ is denoted $E(\boldsymbol{\eta_{k}})$ and is given by

\begin{equation}\label{equaEff}
    E(\boldsymbol{\eta_{k}}) = \sum_{n=1}^{N_k} \eta_{k}^n I_k^n R_k^n.
\end{equation}

Equation (\ref{equaEff}) reflects the relevance of the selected triplets and their ability to be recovered correctly at the reception. 


In the next section, we formulate this objective as an optimization problem and set the constraints for a successful semantic communication.

\section{Problem Formulation}

In resource-constrained networks, it is crucial to select the most relevant
semantic features considering criteria such as the quality of the channel, the transmission time, and the accuracy of recovery. Additionally, in such limited resource networks, when multiple devices transmit large volumes of data simultaneously, it can lead to network congestion, resulting in slower data transmissions and potential communication failures. Therefore, in addition to optimizing network resources and sub-graph selection, it is also important to carefully select devices to achieve the highest transmission efficiency. Hence, our optimization problem is formulated as follows.

\begin{maxi!}|s|
{\boldsymbol{\eta},\boldsymbol\alpha,\boldsymbol{b}, \boldsymbol{P}}{ \sum_{k\in \mathcal{K}}\alpha_k\,  E(\boldsymbol{\eta_k) }\label{mainobj} }
{\label{GeneralOptimizationPb}}{}
\addConstraint{\alpha_kT_k({\boldsymbol{\eta}_k},b_k,P_k)\leq  T^{\rm th}, \quad \forall k\in \mathcal{K} \label{timecons}}
{}{}
\addConstraint{\alpha_kf(\gamma_k(P_k))\leq  \beta^{\rm th}, \quad \forall k\in \mathcal{K} \label{BERcons}}
{}{}
\addConstraint{ 0\leq b_{k}  ,\quad \forall k\in \mathcal{K}, \quad \sum_{k=1}^{K} b_k \leq 1 \label{banwidtth}}
{}{}
\addConstraint{0\leq P_{k} \leq P_k^{\rm max} ,\quad \forall k\in \mathcal{K} \label{powww}}{}{}
\addConstraint{\alpha_{k} \in \{0,1\} ,\quad \forall k\in \mathcal{K}  \label{alphacons}}{}{}
\addConstraint{\eta^n_{k} \in \{0,1\} ,\quad\forall (\!k,n\!)\!\in\!\mathcal{K}\!\!\times\!\! \mathcal{N} \label{etacons}}{}
\end{maxi!}
where $\boldsymbol{\eta}= [\boldsymbol{\eta_1},...,\boldsymbol{\eta_k},...,\boldsymbol{\eta_K}]^T $ , $\boldsymbol{\eta_k} = [\eta_k^1,...,\eta_k^n,...,\eta_k^{N_k}]$, $\boldsymbol{\alpha} = [\alpha_1,...,\alpha_k]$ is the user selection vector, $\boldsymbol{b} = [b_1,...,b_k]$ and $\boldsymbol{P} = [P_1,...,P_k]$. Constraint (\ref{timecons}) ensures that each device $k$ transmits the semantic information within a maximum transmission time $T^{th}$. Constraint (\ref{BERcons}) guarantees that the BER of user $k$ does not exceed a threshold $ \beta^{th}$. Constraints (\ref{banwidtth}) and (\ref{powww}) set the upper limits for the power and bandwidth values.  
Finally, constraints (\ref{alphacons}) and (\ref{etacons}) describe the binary nature of the sub-graph and device selection vectors.

Solving problem (\ref{GeneralOptimizationPb}) directly presents significant challenges for two main reasons. First, the studied optimization falls into the category of mixed integer non-linear programming (MINLP), a class known for its inherent difficulty to solve. Second, even after relaxing the binary constraints, the problem remains complex due to the non-convexity of the objective function and the constraints (\ref{timecons}) and (\ref{BERcons}). 

In the following, we first solve our problem by optimizing the transmit power and bandwidth. However, even with a feasible solution of the power and bandwidth vectors, the problem remains challenging due to its non-convexity. To address this challenge, we introduce an efficient solution that alternates between the sub-graph and device selection.

\section{Joint Optimization of Power, Bandwidth, Sub-graph and User Selection}

In this section, we begin by addressing our problem through the optimization of transmit power and bandwidth. Once we have a feasible solution for the power and bandwidth, the problem reduces into an optimization of sub-graph and device selection that we solve using an iterative algorithm.

\subsection{Power and Bandwidth Allocation}

We first notice that the transmit power and bandwidth vectors are not involved in the objective function. Hence, it is sufficient to set the power and bandwidth vectors such that constraints (\ref{BERcons}) and (\ref{banwidtth}) are active (i.e, the inequalities becomes equalities). Thus, from the binary nature of $\alpha_k$, (\ref{BERcons}) is equivalent to
\begin{equation}
    P_k = \frac{\alpha_kC^{th}}{D_K}, \text{ for } k \in \{1,\dots,K\}, 
    \label{newp}
\end{equation}
with $C^{th} = f^{-1}(\beta^{th})$, and $f^{-1}(.)$ the inverse of the BER function $f(.)$.

Moreover, by substituting equation (\ref{newp}) in  (\ref{timecons}), we get 
\begin{equation}
b_k=\frac{\alpha_k \sum_{n=1}^{N}\eta_k^nS_k^n}{T^{\rm th}B\log_2(1+\alpha_kC^{\rm th})},\; k \in \{1,\dots,K\}.
    \label{newb}
\end{equation}
We note that $b_k$ is well defined when $\alpha_k = 0$ and is equal to $ \lim_{\alpha_k\to0} b_k=\frac{ln(2)\sum_{n=1}^{N}\eta_k^nS_k^n}{T^{\rm th}BC^{\rm th}}$. By replacing the variables $b_k$ and $P_k$ in constraints (\ref{banwidtth}) and (\ref{powww}) with their values, as described by equations (\ref{newp}) and (\ref{newb}) the optimization problem (8) is equivalent to
\begin{maxi!}|s|
{\boldsymbol{\eta},\boldsymbol\alpha}{ \sum_{k\in \mathcal{K}}\alpha_k\, E(\boldsymbol\eta_k) \label{trans_submainobj}}
{\label{trans_subGeneralOptimizationPb}}
{}{}
\addConstraint{\sum_{k=1}^{K} \frac{\alpha_k \sum_{n=1}^{N}\eta_k^nS_k^n}{T^{\rm th}B\log_2(1+\alpha_kC^{\rm th})} \leq 1  \label{sumband1}}
{}{} 
\addConstraint{0\leq \frac{\alpha_kC^{\rm th}}{D_k} \leq P_k^{\rm max} ,\quad \forall k\in \mathcal{K} \label{subpow}}{}{}
\addConstraint{\alpha_{k} \in \{0,1\} ,\quad \forall k\in \mathcal{K}  \label{subalpha}}{}
\addConstraint{\eta^n_{k} \in \{0,1\} ,\quad\forall (\!k,n\!)\!\in\!\mathcal{K}\!\!\times\!\! \mathcal{N}.   \label{subett}}{}
\end{maxi!}

 Given the  power and bandwidth as described by (\ref{newp}) and (\ref{newb}), the problem is reduced to sub-graph and device selection optimization. 
The reformulated problem remains challenging due to its non-convex and binary nature. To solve the studied optimization, we introduce an efficient solution that involves alternating between the sub-graph and device selection.

\subsection{Iterative User and Sub-graph Selection Algorithm}

We propose an algorithm that iteratively solves problem (\ref{trans_subGeneralOptimizationPb}) through optimizing two subproblems, i.e., sub-graph selection subproblem and user selection
subproblem. In particular, at each iteration, we first solve $\boldsymbol{\alpha}$ for a fixed $\boldsymbol{\eta}$, then the optimum $\boldsymbol{\eta}$ is updated based on the obtained value of $\boldsymbol{\alpha}$. \par
First, given $\boldsymbol{\eta}$, problem (\ref{trans_subGeneralOptimizationPb}) becomes
\begin{maxi!}|s|
{\boldsymbol\alpha}{ \sum_{k\in \mathcal{K}}\alpha_k\, E(\boldsymbol\eta_k) \label{submainobj_1}}
{\label{trans_subGeneralOptimizationPb_1}}
{}{}
\addConstraint{\text{Constraints} (\ref{sumband1}),(\ref{subpow}),(\ref{subalpha})}
\end{maxi!}
The challenge of solving problem (\ref{trans_subGeneralOptimizationPb_1}) stems from the non-convexity of sum-ratio constraint (\ref{sumband1}). To tackle this constraint, a local convex approximation is applied.
Specifically, by noticing that $\log_2(1+\alpha_kC^{\rm th}) \geq \frac{\alpha_kC^{\rm th}}{ln(2)(\alpha_kC^{\rm th}+1)}$ for a device $k$, we obtain 
     \begin{equation}
         \sum_{k=1}^{K} \frac{\alpha_k \sum_{n=1}^{N}\eta_k^nS_k^n}{T^{\rm th}B\log_2(1+\alpha_kC^{\rm th})}\leq \frac{ln(2)}{T^{\rm th}B}\!\!\sum_{k=1}^{K} (\alpha_k+ \frac{1}{C^{\rm th}})\sum_{n=1}^{N}\eta_k^nS_k^n.
         \label{eqlog_ineq}
     \end{equation} 
We define the new constraint as follows 
\begin{equation}
    \frac{ln(2)}{T^{\rm th}B}\!\!\sum_{k=1}^{K} (\alpha_k+ \frac{1}{C^{\rm th}})\sum_{n=1}^{N}\eta_k^nS_k^n \leq 1,
    \label{cvxcst}
\end{equation}
which is convex since the left-hand-side (LHS) is linear with respect to $\alpha_k$. Fortunately, this convex relaxation results in a linear problem with respect to $\boldsymbol{\alpha}$ which can be efficiently solved using a linear programming algorithm such as the one described in \cite{linear2021comp}.

Inequality (\ref{eqlog_ineq}) shows that the convex constraint in (\ref{cvxcst}) always implies the non-convex constraint (\ref{sumband1}). 
However, solving the user selection problem using constraint (\ref{cvxcst}) instead of (\ref{sumband1}) may result in a sub-optimal value of the objective function. In fact, there may be an $\boldsymbol{\alpha}$ that does not satisfy inequality (\ref{cvxcst}) and gives a better solution to problem (\ref{trans_subGeneralOptimizationPb_1}). To address this problem, we
track the difference between the two constraints while solving problem (\ref{trans_subGeneralOptimizationPb_1}) with constraint (\ref{cvxcst}) instead of (\ref{sumband1}).

In fact, we first measure the gap between the non-convex and convex constraints. We denote this gap as $\Delta$, which is given by

\begin{equation}
    \begin{split}
        \Delta&=\! \sum_{k=1}^{K} \frac{\sum_{n=1}^{N}\eta_k^nS_k^n}{T^{\rm th}B}\!\left(ln(2)(\alpha_k\!+\! \frac{1}{C^{\rm th}})\!-\!(\frac{\alpha_k}{\log_2(1\!+\!\alpha_kC^{\rm th})}\!) \!   \right)\\
        &  \stackrel{(a)}\leq \sum_{k=1}^{K} \frac{\sum_{n=1}^{N}\eta_k^nS_k^n}{T^{\rm th}B} \left(ln(2)\alpha_k + \frac{ln(2) - ln(2)}{C^{th}}\right) \\
        &\stackrel{(b)}\leq \sum_{k=1}^{K} \frac{\sum_{n=1}^{N}ln(2)S_k^n}{T^{\rm th}B}\\
        & 
\triangleq \Delta_{max},
     \end{split}
\end{equation}
where (a) comes from the fact that $\log_2(1+\alpha_kC^{\rm th}) \leq \frac{\alpha_kC^{\rm th}}{ln(2)}$. Furthermore, (a) takes a maximum upper bounded in $\alpha_k = 1$ and $\eta_k^n = 1$, hence, the approximation in (b). To reduce the gap between the two constraints, we define a new upper bound to (\ref{cvxcst}) by adding a correction term of $\Delta_b=\frac{\Delta_{max}}{2}$ to the current upper bound. Consequently, constraint (\ref{cvxcst}) becomes 

\begin{equation}
    \frac{ln(2)}{T^{\rm th}B}\!\!\sum_{k=1}^{K} (\alpha_k+ \frac{1}{C^{\rm th}})\sum_{n=1}^{N}\eta_k^nS_k^n \leq 1+\Delta_b.
    \label{cvxcst2}
\end{equation}

The user selection problem is initially addressed by applying constraint (\ref{cvxcst2}) with the adjusted upper bound. Subsequently, we verify whether the resulting vector $\boldsymbol{\alpha}$ satisfies the original constraint (\ref{sumband1}). If it does, the corrected term can be further incremented by $\frac{\Delta_{max}}{4}$, and the problem is solved once more. Conversely, if the original constraint is not met, the corrected term is reduced by $\frac{\Delta_{max}}{4}$, and the problem is solved. This iterative process continues until a desired level of accuracy $\epsilon_2$ is attained. Lines $3$ to $23$ of Algorithm \ref{itera_modified} describe this process.




After successfully addressing the user selection problem, we solve the sub-graph selection which is formulated as follows
\begin{maxi!}|s|
{\boldsymbol\eta}{ \sum_{k\in \mathcal{K}}\alpha_k\, E(\boldsymbol\eta_k) \label{submainobj_1}}
{\label{trans_subGeneralOptimizationPb_2}}
{}{}
\addConstraint{\text{Constraints} (\ref{sumband1}),(\ref{subett})}
\end{maxi!}
Problem (\ref{trans_subGeneralOptimizationPb_2}) is linear and can be efficiently solved using a linear programming algorithm \cite{linear2021comp}.
\begin{algorithm}
	\caption{Semantic Information and User Selection} 
	\label{itera_modified}
     \hspace*{\algorithmicindent} \textbf{Input:} {A feasible solution ({\bfseries $\alpha$}$^0$, {\bfseries $\eta$}$^0$), $\epsilon_1, \epsilon_2 > 0 $ and iteration number $\ell$ = 0}
	\begin{algorithmic}[1]
  \Repeat 
    \State With given $\boldsymbol{\eta}^\ell$,
    \State  $\Delta_b = \frac{\Delta_{max}}{2}$
    \State $bound = 1 + \Delta_b$
    \While {$\Delta_b \geq \epsilon_2$}
        \State {Solve the problem in (\ref{trans_subGeneralOptimizationPb_1}) using constraint (\ref{cvxcst2}) with the defined $bound$ by linear programming algorithms and obtain $\boldsymbol{\alpha}^{\ell+1}$}
        \State {$\Delta_b = \frac{\Delta_b}{2}$}
        \If{ ($\boldsymbol{\alpha}^{\ell+1}$, $\boldsymbol{\eta}^{\ell}$) satisfies constraint (\ref{sumband1})}
        \State {$bound = bound + \Delta_b$}
        \Else 
        \State{$bound = bound - \Delta_b$}
        \EndIf
    \EndWhile
\State With given $\boldsymbol{\alpha}^{\ell+1}$, solve the problem in (\ref{trans_subGeneralOptimizationPb_2}) and obtain the solution $\boldsymbol{\eta}^{\ell+1}$ using linear programming algorithms
    \State Set $\ell$ = $\ell$+1
    \Until{$|\sum_{k\in \mathcal{K}}\alpha_k^{\ell+1}\, A(\eta_k^{\ell+1}) - \sum_{k\in \mathcal{K}}\alpha_k^{\ell}\, A(\eta_k^{\ell})| \leq \epsilon_1 $}
	\end{algorithmic} 
\end{algorithm}

Finally, Algorithm \ref{itera_modified} gives the solution of the optimization problem stated in (\ref{trans_subGeneralOptimizationPb}) by iteratively solving problems in (\ref{trans_subGeneralOptimizationPb_1}) and (\ref{trans_subGeneralOptimizationPb_2}). At each iteration, the optimal solutions of (\ref{trans_subGeneralOptimizationPb_1}) and (\ref{trans_subGeneralOptimizationPb_2}) are obtained and thus, the objective in (\ref{trans_subGeneralOptimizationPb}) is increased. Furthermore, since the user allocation and semantic information allocation are upper-bounded by $1$, the objective is also upper-bounded. Therefore, the convergence of the algorithm to a local optimum is guaranteed.
\subsection{Complexity analysis}
To solve problem (\ref{trans_subGeneralOptimizationPb}) using Algorithm \ref{itera_modified}, the  complexity depends on the number of iterations $\mathcal{L}$ of the iterative maximization Algorithm \ref{itera_modified}, and on the complexity needed to solve each of the two subproblems. As for problem (\ref{trans_subGeneralOptimizationPb_1}) with corrected bound, a complexity of $\mathcal{O}(\log_2(\Delta_{max}))$ is needed to find the optimal $bound$ and the optimal $\boldsymbol{\alpha}$ is obtained with a computational complexity $\mathcal{O}(N^w)$, where $w \leq 2.5$, \cite{linear2021comp}. On the other hand, the optimization of $\boldsymbol{\eta}$ in problem (\ref{trans_subGeneralOptimizationPb_2}) involves a complexity of $\mathcal{O}((N\times M)^w)$. As a result, the total complexity of Algorithm \ref{itera_modified} can be evaluated as $\mathcal{O}\left(\mathcal{L}N^w(\log_2(\Delta_{max}) + M^w)\right)$.
\section{Simulation Results}

To assess the performance of our approach, we consider an area of $1km^2$ where $K = 10$ devices are scattered randomly and communicate with a BS in the middle
of the area. We assume a total bandwidth of $B = 1M Hz$ and a BER threshold of $\beta^{th} = 10^{-5}$. The power spectral
density of the Gaussian noise is equal to $\sigma^2 = 10^{-14}mW/Hz$. 
 We use the graph generation model in \cite{REB2021-rebel-relation} for semantic extraction and the python \textit{SentenceTransformers} for data embedding to vectorize the semantic information. We also use web pages generated from Google as a dataset. We suppose that the data size of an English letter is $8$ bits, thus each triplet $n$ of device $k$ is encoded by $S_k^n = E_n\times 8$ with $E_n$ is the number of letters in the triplet $n$. We consider a $100\% $ SE when all the devices are participating in the communication and are sending all their information. We also assume that the BER function $f$ is of the form of  $f(\gamma(P_k)) = \frac{1}{\gamma(P_k)}$. Unless stated otherwise, $P_{max} = 0.01W$ and $T^{th}= 8 ms$.\par

\begin{figure}[htp] 
    \centering
    \subfloat[Context = "Girl description"]{%
        \includegraphics[width=0.20\textwidth]{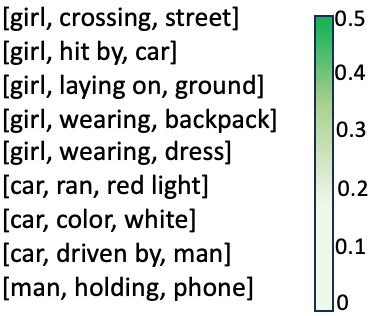}%
        \label{fig:a}%
        }%
    \hfill%
    \subfloat[Context: "Car information"]{%
        \includegraphics[width=0.20\textwidth]{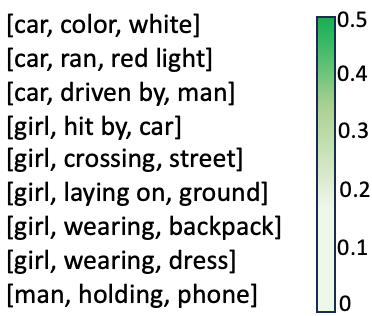}%
        \label{fig:b}%
        }%
    \caption{Importance score per triplet for different contexts}
    \label{Importance score}
\end{figure}

In Figure \ref{Importance score}, we present an example of triplets generated from a web page describing an accident. We assign a color to the importance score of each triplet. In particular, as the important score increases, the color changes from white to green. We can observe that the same triplets have different importance scores depending on the context. For example, in a context where the required information is about a car involved in the accident, the triplets describing the car have more importance than the others. \par 
  \begin{figure} 
\centering
{%
  \includegraphics[width=0.38\textwidth]{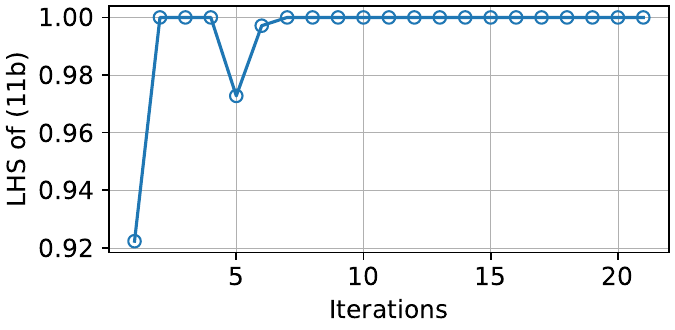}%
  }
\caption{Value of LHS of constraint (11b)}
\label{newbound}
\end{figure}

In Figure \ref{newbound}, we compute the LHS of constraint (\ref{sumband1}) using the optimal parameters resulted from our algorithm per iteration with an accuracy $\epsilon_2 = 10^{-5}$. This shows that the convex approximation constraint we defined approaches the initial non convex one which guarantees an efficient solution to our problem.
\par

  \begin{figure}
\centering
{%
  \includegraphics[width=0.38\textwidth]{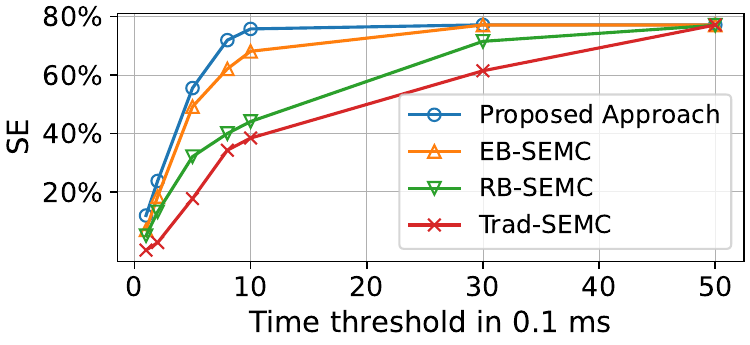}%
  }
\caption{SE per time threshold for different scenarios}
\label{Accu_per_scenarios}
\end{figure}
 Figure \ref{Accu_per_scenarios} plots the SE for different semantic-aware scenarios: i) EB-SEMC that uses the proposed optimization with fixed equal bandwidth ; ii) RB-SEMC where bandwidth is randomly allocated across devices ; iii) Trad-SEMC where users transmit their semantic information bit by bit until constraints of the problem are reached i,e,. time threshold $T^{th}$ and BER threshold $B^{th}$. The SE of all schemes increases with time threshold. This is because large time threshold allows more transmission of data. Our method outperforms the other benchmarks since it allocates the optimal bandwidth while it is fixed for EB-SEMC and random for RB-SEMC. Furthermore, Trad-SEMC is the least efficient as the latter does not take into consideration the relevance of information while transmission.
 
\begin{figure}
\centering
{%
  \includegraphics[width=0.38\textwidth]{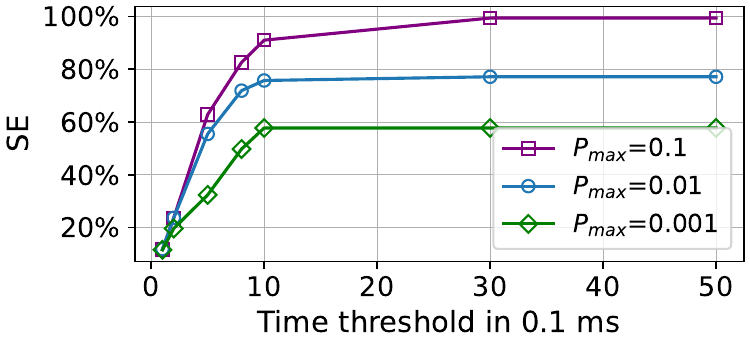}%
  }
\caption{SE per time threshold for different $P_{max}$}
\label{Accu_per_power}
\end{figure}

 Figure \ref{Accu_per_power} shows the SE of our approach per time threshold for different $P_{max}$. In our setup, a $100\%$ SE is achieved when $P_{max}$ is equal to $0.1 W$ and the time threshold is  $T^{th}=3ms$. Depending on time and power constraints, the proposed approach selects the optimal devices and the most relevant information to achieve the highest efficiency. As can be seen from the figure, as the power increases, less transmission time is required to achieve a higher semantic efficiency.\par

\section{Conclusion}
In this work, we tackle semantic communication where limited number of devices operate under wireless communications constraints. We formulate an optimization problem that aims to select the most relevant semantic information over devices while satisfying the resource constraints.
We present  an efficient  iterative algorithm to resolve our problem with guarantee of convergence and perform numerical experiments to validate the efficiency of our framework.
 \balance
\bibliographystyle{IEEEbib}
\bibliography{references}

\end{document}